\documentclass[12pt,epsf]{article}
\usepackage{graphicx}
\begin{document}

\def\a{\alpha}
\def\b{\beta}
\def\c{\varepsilon}
\def\d{\delta}
\def\e{\epsilon}
\def\f{\phi}
\def\g{\gamma}
\def\h{\theta}
\def\k{\kappa}
\def\l{\lambda}
\def\m{\mu}
\def\n{\nu}
\def\p{\psi}
\def\q{\partial}
\def\r{\rho}
\def\s{\sigma}
\def\t{\tau}
\def\u{\upsilon}
\def\v{\varphi}
\def\w{\omega}
\def\x{\xi}
\def\y{\eta}
\def\z{\zeta}
\def\D{\Delta}
\def\G{\Gamma}
\def\H{\Theta}
\def\L{\Lambda}
\def\F{\Phi}
\def\P{\Psi}
\def\S{\Sigma}

\def\o{\over}
\def\beq{\begin{eqnarray}}
\def\eeq{\end{eqnarray}}
\newcommand{\gsim}{ \mathop{}_{\textstyle \sim}^{\textstyle >} }
\newcommand{\lsim}{ \mathop{}_{\textstyle \sim}^{\textstyle <} }
\newcommand{\vev}[1]{ \left\langle {#1} \right\rangle }
\newcommand{\bra}[1]{ \langle {#1} | }
\newcommand{\ket}[1]{ | {#1} \rangle }
\newcommand{\EV}{ {\rm eV} }
\newcommand{\KEV}{ {\rm keV} }
\newcommand{\MEV}{ {\rm MeV} }
\newcommand{\GEV}{ {\rm GeV} }
\newcommand{\TEV}{ {\rm TeV} }
\def\diag{\mathop{\rm diag}\nolimits}
\def\Spin{\mathop{\rm Spin}}
\def\SO{\mathop{\rm SO}}
\def\O{\mathop{\rm O}}
\def\SU{\mathop{\rm SU}}
\def\U{\mathop{\rm U}}
\def\Sp{\mathop{\rm Sp}}
\def\SL{\mathop{\rm SL}}
\def\tr{\mathop{\rm tr}}

\def\IJMP{Int.~J.~Mod.~Phys. }
\def\MPL{Mod.~Phys.~Lett. }
\def\NP{Nucl.~Phys. }
\def\PL{Phys.~Lett. }
\def\PR{Phys.~Rev. }
\def\PRL{Phys.~Rev.~Lett. }
\def\PTP{Prog.~Theor.~Phys. }
\def\ZP{Z.~Phys. }


\baselineskip 0.7cm

\begin{titlepage}

\begin{flushright}
UT--10--17\\
IPMU--10--0158\\
\end{flushright}

\vskip 1.35cm
\begin{center}
{\large \bf
Split Generation in the SUSY Mass Spectrum \\ and $B_s-{\bar B}_s$ Mixing
}
\vskip 1.2cm
Motoi Endo, Satoshi Shirai and Tsutomu T. Yanagida
\vskip 0.4cm

{\it  Department of Physics, University of Tokyo,
Tokyo 113-0033, Japan\\
Institute for the Physics and Mathematics of the Universe, 
University of Tokyo,\\ Chiba 277-8568, Japan\\}

\vskip 1.5cm

\abstract{
We show that the like-sign di-muon anomaly reported recently by the D0 Collaboration can 
be explained in the supersymmetric standard model (SM) if the squarks and the sleptons in 
the first two generations have relatively small, but degenerate mass spectrum, and those in 
the third generation are larger as $O(1-10)$TeV. This split generation model provides large
contributions to the $B_s-{\bar B}_s$ mixing, although most of the FCNC's are suppressed 
due to the large masses of the third generation squarks or the GIM mechanism partially 
acting on the first and second generations.

}
\end{center}
\end{titlepage}

\setcounter{page}{2}

\section{Introduction}

Gravity mediation~\cite{Nilles:1983ge} is a well-motivated and simple mechanism for transmitting 
supersymmetry (SUSY) breaking 
to the observed sector. However, it predicts too much flavor-changing neutral currents (FCNC's) in a generic vacuum of the gravity mediation. The most popular solution to this FCNC
problem is to postulate a degenerate SUSY-breaking  mass spectrum for squarks and sleptons. 
However, the origin of such a mass degeneracy  is not known. An alternative solution was
considered where the masses of squarks and sleptons in the first and second generations are 
relatively large as $O(10)$TeV and the masses for the third generation are in the range of 
$O(100)$GeV$-1$TeV. This is called as ``decoupling model"~\cite{decoupling}.

Very recently, a new solution to the FCNC problem has been proposed where the squarks and sleptons 
in the first and second generations have relatively small, but degenerate SUSY-breaking masses 
and the squarks and sleptons in the third generation are larger as $O(1-10)$TeV. 
This is motivated by the Nambu-Goldstone (NG) hypothesis for quarks and leptons 
in the first two generations~\cite{MNSY}.
In fact, the SUSY non-linear sigma model,  $E_7/SO(10)\times U(1)^2$, is known~\cite{KY,YY} 
to accommodate two generations of quark and lepton chiral multiplets as NG modes~\cite{BPY}.
We call it ``split generation model" to distinguish it from the decoupling model

We show, in this paper, that the split generation model may explain the like-sign di-muon 
anomaly observed recently in the D0 experiment~\cite{Abazov:2010hv}, without generating any 
conflict with all experimental constraints. 
This is because it naturally predict a relatively large $B_s-{\bar B}_s$ mixing with a new CP phase,
while other FCNC processes are suppressed. 
In particular, it is stressed that the model can easily satisfy the bound from 
${\rm Br}(b \to s\gamma)$,  in contrast to other two models with the degenerate or the decoupling mass spectra. This is the main reason why we propose the split generation model as a serious candidate
for the beyond SM.

Before going to detailed analyses in this paper, it should be discussed here on the other
problem in the gravity mediation, that is, the $CP$ problem. The gravity mediation involves several
$CP$ violating phases in SUSY-breaking soft masses and it predicts too large $CP$ violation 
at low energies.
Since there has been found no solution to this problem so far, we simply assume, throughout this paper, 
that $CP$ is not broken in the SUSY breaking sector. Therefore, all $CP$ violating phases
originate only from the diagonalization of quark and lepton mass matrices. From the theoretical point of view, 
it is more interesting to impose $CP$ invariance in the full theory, and the CP-violating phase in the 
CKM matrix appears as a result of some spontaneous CP violation. This assumption of $CP$ invariance itself 
may give a deep clue to a solution of the strong $CP$ problem, since the QCD vacuum angle $\theta_{\rm QCD}$ vanishes
at the fundamental cut-off scale. 
However, the quark rotations for the diagonalization of their mass matrices, in general, shift 
the QCD $\theta$ angle and hence we have the strong $CP$ violation even if the original QCD vacuum is $CP$ invariant. 
Therefore, the hypothesis of $CP$ invariance for the full theory does not
solve the strong $CP$ problem at the first glance. However, if the quark mass matrices are hermitian, the
shift of the $\theta_{\rm QCD}$ angle vanishes and we can maintain $\theta_{\rm QCD}=0$ even after the 
diagonalization of quark
mass matrices~\cite{Hiller:2002um}. This assumption on quark mass matrices becomes also very important to suppress neutron and atomic electric dipole moments sufficiently, as shown in section 5.

\section{Split Generation Model}

We consider the split generation model, in which the squarks have the following spectrum of 
SUSY-breaking soft masses,
\begin{eqnarray}
 \tilde m_q = {\rm diag}\,(m_{\tilde q}, m_{\tilde q}, M_{\tilde q}),
 \label{eq:squark-mass}
\end{eqnarray}
for the three generations. Here,``split generation" means that the squarks in the first two generations 
have relatively small and degenerate SUSY-breaking soft masses, $m_{\tilde q}$, and the masses 
in the third generation are relatively large, $M_{\tilde q} > m_{\tilde q}$.
On the other hand, the Yukawa matrices of the quarks are assumed to be non-diagonal. 
We call this basis as the generation basis. 
\footnote{Note that this does not define the generation axes uniquely, since the soft masses of the 
first and second generations are degenerate, and we can rotate their axes without changing the 
SUSY-breaking soft mass matrices (\ref{eq:squark-mass}). Here, we introduce (very) tiny violations 
of the mass degeneracy, though they 
are neglected hereafter, since they are irrelevant for the following discussion. They are needed 
only for defining the generation basis. }

The model generally induces FCNC's and CP violations through the mixings with the third generation. 
In the super-CKM basis, where the SUSY-invariant Yukawa matrices of the quark chiral multiplets 
are diagonalized, not only quarks but also squarks are rotated from the generation basis, and 
hence mass matrices of squarks are no longer diagonal due to the non-degenerate third component 
of (\ref{eq:squark-mass}). Moreover, since the Yukawa matrices are generally complex, the squark 
mass matrices acquire complex phases in the super-CKM basis independent of the CKM phase. 
\footnote{ The CP violations are also represented by the Jarlskog invariants in SUSY~\cite{Lebedev:2002wq}.}

In order to analyze the (CP-violating) FCNC's quantitatively, it is convenient to use the hybrid 
basis rather than the super-CKM basis, where only the quark mass matrices are diagonalized, 
while the squarks are left unrotated. In the hybrid basis, the gaugino--quark--squark vertices 
are flavor non-diagonal. Actually, when we change the basis from the generation to the hybrid 
basis, the gluino vertices are transformed as 
\begin{eqnarray}
 \mathcal{L} = -\sqrt{2} g_s \bar q_L \tilde g^{(a)} T^{(a)} \tilde q_L + \cdots
 = -\sqrt{2} g_s \bar q'_L (U_{L})^\dagger \tilde g^{(a)} T^{(a)} \tilde q_L + \cdots,
\end{eqnarray}
where $g_s$ is the QCD gauge coupling constant with the color index $(a)$, and the unitarity 
matrices $(U_{L(R)})_{ij}$ are diagonalization matrices for the mass matrices of the left- (right-) 
handed quarks. Since the rotation matrices of the up- and down-type quarks provide the CKM 
matrix $V$, the mixing angles $(U_{L,R})_{23}$ and $(U_{L,R})_{32}$ are expected to be $\sim 
V_{ts} \sim \lambda^2$ with the Wolfenstein parameter $\lambda \simeq 0.2$. 

The squark couplings to the Higgs/Higgsino can also change flavors. In particular, the chirality 
mixing of the squark mass matrix, $(\tilde m^q_{LR,RL})^2$, have flavor off-diagonal elements. 
In fact, it depends on the Yukawa coupling such as $(\tilde m^d_{LR,RL})^2_{ij} \simeq -(Y_d)_{ij} 
v \mu \tan\beta$ for the down-type squarks, where $Y_d$ is the down-type Yukawa coupling, 
$v \simeq 174$GeV the SM Higgs vacuum expectation value (VEV), $\mu$ 
the Higgsino mass, and $\tan\beta$ a ratio of the up- and down-type Higgs VEVs. It is noticed 
that $Y_q$ in $(\tilde m^q_{LR,RL})^2$ is non-diagonal in the hybrid basis and is represented 
as $Y_q = U_L Y_q^{\rm (diag)} U_R^\dagger$, where $Y^{\rm (diag)}$ is the diagonalized 
Yukawa coupling. Similarly, the squark couplings to the neutralino/chargino have flavor-changing 
components, which come from the mixing between the neutralino/chargino and the Higgsino.

\section{$B_s - \bar B_s$ Oscillation}

We first discuss the oscillation between the $B_s^0$ and $\bar B_s^0$ mesons. 
In fact, the like-sign di-muon asymmetry is sensitive to the oscillation. 
The D0 Collaboration recently reported the asymmetry~\cite{Abazov:2010hv},
\begin{eqnarray}
 A_{\rm sl}^b \equiv \frac{N_b^{++} - N_b^{--}}{N_b^{++} + N_b^{--}} 
 = -(9.57 \pm 2.51 \pm 1.46) \times 10^{-3},
\end{eqnarray}
where $N_b^{++(--)}$ is the number of $b\bar b \to \mu^{+(-)}\mu^{+(-)} X$ events. This 
result is 3.2$\sigma$ deviated from the SM prediction, $A_{\rm sl}^b = 
(-2.3^{+0.5}_{-0.6}) \times 10^{-4}$~\cite{Lenz:2006hd}. At the Tevatron, it is related to 
the semileptonic CP asymmetries of the $B_d$ and $B_s$ mesons~\cite{Abazov:2010hv} as,
\begin{eqnarray}
 A_{\rm sl}^s = (0.506 \pm 0.043) a_{\rm sl}^d + (0.494 \pm 0.043) a_{\rm sl}^s,
\end{eqnarray}
where $a_{\rm sl}^d$ and $a_{\rm sl}^s$ are given as
\begin{eqnarray}
 a_{\rm sl}^q = {\rm Im} \frac{\Gamma_{12}^q}{M_{12}^q} 
 = \left|\frac{\Gamma_{12}^q}{M_{12}^q}\right| \sin\phi^q,
\end{eqnarray}
in terms of the off-diagonal elements of the mass and decay matrices, $M_{12}^q$ and 
$\Gamma_{12}^q$, of the $B_q - \bar B_q$ oscillation for $q = d, s$. 

The oscillation contributes to other observables as well as the like-sign di-muon asymmetry 
(see \cite{Ligeti:2010ia} and \cite{Bauer:2010dg}). 
In particular, the mass difference of the $B_q$ mesons, $\Delta m_q$, provides a severe 
bound on the SUSY contributions. Additionally, the time-dependent $B_s \to \psi\phi$ decay 
give a measurement of the width difference of the $B_s$ mesons, $\Delta\Gamma_s$, 
and the time-dependent CP asymmetry, $S_{\psi\phi}$. Combining these observables, we 
obtain the $\chi^2$ minimum at~\cite{Ligeti:2010ia,Bauer:2010dg}
\footnote{
 After the analysis in \cite{Ligeti:2010ia,Bauer:2010dg}, the D0 Collaboration updated 
 the result of $B_s \to \psi\phi$~\cite{D0-psiphi}. Although it turned to be SM consistent, 
 the minimum positions are expected to be unaffected so much.
}
\begin{eqnarray}
 (h_s, \sigma_s) \simeq (0.5, 120^\circ)\ {\rm and}\ (1.8, 100^\circ),
 \label{eq:dimuon}
\end{eqnarray}
for the $B_s$ meson, where $h_s$ and $\sigma_s$ are defined by 
$M_{12}^s = (M_{12}^s)^{\rm SM} (1 + h_s e^{2i\sigma_s})$. 
Actually, this significantly contributes to the phase, $\phi^q$, away from the SM value in order 
to explain the like-sign di-muon anomaly. 
On the other hand, the contribution to the $B_d$ meson is constrained to be small.

The split generation model dominantly contributes to the $B_s - \bar B_s$ oscillation 
as is shown below. 
The oscillation is represented by the following $\Delta B = 2$ effective Hamiltonian,
\begin{eqnarray}
 H_{\rm eff} = \sum_{i=1}^5 C_i O_i + \sum_{i=1}^3 \tilde C_i \tilde O_i,
\end{eqnarray}
where the operators are
\begin{eqnarray}
 && O_1 = (\bar s^\alpha_L \gamma_\mu b^\alpha_L) 
 (\bar s^\beta_L \gamma^\mu b^\beta_L), \nonumber\\
 && O_2 = (\bar s^\alpha_R b^\alpha_L) (\bar s^\beta_R b^\beta_L),~~~
 O_3 = (\bar s^\alpha_R b^\beta_L) (\bar s^\beta_R b^\alpha_L), \nonumber\\
 && O_4 = (\bar s^\alpha_R b^\alpha_L) (\bar s^\beta_L b^\beta_R),~~~
 O_5 = (\bar s^\alpha_R b^\beta_L) (\bar s^\beta_L b^\alpha_R),
 \label{eq:BB-operator}
\end{eqnarray}
and $\tilde O_i$ by $R \leftrightarrow L$. In the decoupling limit, $M_{\tilde q} \gg m_{\tilde q}$, 
the SUSY contributions to the Wilson coefficients become
\begin{eqnarray}
 && C_1 = -\frac{\alpha_s^2}{216} \frac{1}{m_{\tilde q}^2}
 \left[ 24 x f(x) + 66 \tilde f(x) \right] (U_{L})_{23}^2, 
 \label{eq:Wilson1}\\
 && C_2 = -\frac{\alpha_s^2}{216} \frac{1}{m_{\tilde q}^2}
 204\, x f(x) (\delta_{RL})_{23}^2,~~~
 C_3 = \frac{\alpha_s^2}{216} \frac{1}{m_{\tilde q}^2}
 36\, x f(x) (\delta_{RL})_{23}^2, \\
 && C_4 = -\frac{\alpha_s^2}{216} \frac{1}{m_{\tilde q}^2}
 \left[(504 x f(x) - 72 \tilde f(x))
   (U_{L})_{23}(U_{R})_{23} 
   -132 \tilde f(x) (\delta_{LR})_{23} (\delta_{RL})_{23} 
 \right], \\
 && C_5 = -\frac{\alpha_s^2}{216} \frac{1}{m_{\tilde q}^2}
 \left[(24 x f(x) + 120 \tilde f(x))
   (U_{L})_{23}(U_{R})_{23} 
   -180 \tilde f(x) (\delta_{LR})_{23} (\delta_{RL})_{23}  
 \right], \\
 && \tilde C_1 = -\frac{\alpha_s^2}{216} \frac{1}{m_{\tilde q}^2}
 \left[24 x f(x) + 66 \tilde f(x)\right] (U_{R})_{23}^2, \\
 && \tilde C_2 = -\frac{\alpha_s^2}{216} \frac{1}{m_{\tilde q}^2}
 204\, x f(x) (\delta_{LR})_{23}^2,~~~
 \tilde C_3 = \frac{\alpha_s^2}{216} \frac{1}{m_{\tilde q}^2}
 36\, x f(x) (\delta_{LR})_{23}^2,
 \label{eq:Wilson2}
\end{eqnarray}
where the LR and RL mixing parameters are defined as $(\delta_{LR,RL})_{ij} \equiv 
(\tilde m^d_{LR,RL})^2_{ij}/M_{\tilde q}^2$, and the loop functions are
\begin{eqnarray}
 && f(x) = \frac{2(1-x)+(1+x)\ln x}{(x-1)^3},~~~ 
 \tilde f(x) = \frac{1-x^2+2x\ln x}{(x-1)^3},
\end{eqnarray}
with a ratio of the gluino mass to the light squark mass, $x = m_{\tilde g}^2/m_{\tilde q}^2$. 
In the expression, the terms with $(U_{L})_{23}$ and $(U_{R})_{23}$ come from the strange 
squark diagrams, while the bottom-squark contributions with $(U_{L})_{23}$ and $(U_{R})_{23}$ 
are discarded, since they are suppressed by a heavy squark mass $M_{\tilde q}$. 
On the other hand, the contributions with the chirality flip, $(\delta_{LR,RL})_{ij}$, are always 
suppressed by $M_{\tilde q}$, because they inevitably depend on the scalar-bottom quark. 

From the above effective Hamiltonian, we obtain the dispersive part of the $B_s - \bar B_s$ 
mixing amplitude, $M_{12}^s \equiv \langle B_s^0 | H_{\rm eff}| \bar B_s^0 \rangle$, in terms of 
the hadron matrix elements, $\langle B_s^0 | O_i | \bar B_s^0 \rangle$. Numerically, it becomes
\begin{eqnarray}
 h_s e^{2i\sigma_s} &=& \frac{(M_{12})^{\rm SUSY}}{(M_{12})^{\rm SM}} 
 \label{eq:b-bbar} \\
 &\simeq&
 \left[
 66 \left[(U_{L})_{23}^2 + (U_{R})_{23}^2\right] - 
 1.4 \times 10^3 (U_{L})_{23}(U_{R})_{23} 
 \right. 
 \nonumber \\
 && 
 \left. + 
 2.1 \times 10^2 \left[(\delta_{RL})_{23}^2 + (\delta_{LR})_{23}^2\right] -
 1.0 \times 10^3 (\delta_{LR})_{23} (\delta_{RL})_{23}
 \right] 
 \left( \frac{m_{\tilde q}}{400{\rm GeV}} \right)^{-2},
 \nonumber
\end{eqnarray}
for $m_{\tilde q} = m_{\tilde g}$. We find that the oscillation is enhanced when 
both $(U_{L})_{23}$ and $(U_{R})_{23}$ are large. Then, the CP symmetry is violated by their 
phases. Although a coefficient of the term depends on the mass spectrum, it can be checked 
that it is less sensitive to $x$. For instance, it becomes $1.4(1.0) \times 10^3$ for 
$m_{\tilde g} = 400(1000)$GeV and $m_{\tilde q} = 400$GeV. 
We also obtain the similar result for the $B_d - \bar B_d$ oscillation, but the mixing is provided by 
$(U_{L, R})_{13}$ and $(\delta_{LR,RL})_{13}$. Thus, it is suppressed.

From (\ref {eq:b-bbar}), we estimate the SUSY contributions to the $B_s - \bar B_s$ oscillation. 
Numerically, we find that they become as large as the value which is required to explain the 
di-muon anomaly (\ref{eq:dimuon}). Actually, postulating $|(U_{L})_{23}| = |(U_{R})_{23}| = 
\lambda^2$ with $\lambda \simeq 0.2$, we obtain $h_s = 1 - 2$ and $\simeq 0.5$ for $m_{\tilde q} 
= 420 - 600$GeV and $\simeq 900$GeV, respectively. On the other hand, the anomalous like-sign 
di-muon charge asymmetry requires a large CP violation in $B_s - \bar B_s$. When the mixings 
are $|(U_{L})_{23}| \simeq |(U_{R})_{23}|$, $\sigma_s$ is related to the phase of $(U_{L})_{23}$ 
and $(U_{R})_{23} $ as 
\begin{eqnarray}
  \Theta^+ \equiv {\rm arg}[(U_{L})_{23}(U_{R})_{23}] \simeq 2\sigma_s - 180^\circ.
  \label{eq:phase+}
\end{eqnarray}
Thus, from (\ref{eq:dimuon}) the phases are required to be $\Theta^+ \simeq 60^\circ$ and $20^\circ$ 
for $h_s \simeq 0.5$ and $1.8$, respectively.

When masses of the third generation squarks are heavy, the $(\delta_{LR,RL})$ contributions 
are very small, since $(\delta_{LR,RL})$ are suppressed by $1/M_{\tilde q}^2$. Actually, taking 
a typical value for the Yukawa coupling, $(Y_d)_{23} \simeq (0.01-0.1)$, we find that they are too 
small to satisfy (\ref{eq:dimuon}) for $M_{\tilde q} \gsim 1$TeV. Thus, we will discard them in the 
following.

There are two comments on the SUSY contributions to the like-sign di-muon anomaly. 
In the split generation model, the SUSY contribution to the $B_d$ oscillation is suppressed. 
This is because the mixings are expected to be $|(U_{L})_{13}| \simeq |(U_{R})_{13}| \simeq 
10^{-(2-3)}$ from the CKM matrix. We obtain $h_d \lsim 0.1$ even for $m_{\tilde q} = 420 - 600$GeV. 
Since the contribution is small, it is difficult to explain the di-muon anomaly solely by the $B_d$ 
mixing, rather this can improve $\chi^2$ with a proper phase when the $B_s$ contribution is 
large~\cite{Ligeti:2010ia,Bauer:2010dg}. Also, the di-muon anomaly favors $\Gamma_{12}^s$ 
to be deviated from the SM prediction, although uncertainties are still 
large~\cite{Ligeti:2010ia,Bauer:2010dg}. However, the split generation model does not change 
$\Gamma_{12}^s$ from the SM value.

As a final remark in this section, we discuss the experimental constraint from the $K-\bar K$ mixing.
The SUSY contribution is obtained by substituting $[(U_{L,R})_{23}(U_{L,R}^*)_{13}]^2$ for 
$(U_{L,R})_{23}^2$ in Eq.~(\ref{eq:Wilson1})--(\ref{eq:Wilson2}). This is because the leading 
contributions which depend on $(U_{L,R})_{12}$ almost cancel out due to the GIM mechanism. 
However, the light down and strange squarks contribute to the $K-\bar K$ mixing through the 
term of  $[(U_{L,R})_{23}(U_{L,R}^*)_{13}]$ (due to the unitarity condition $\sum_i (U_{L(R)})_{2i}
(U_{L(R)}^*)_{1i} = 0$). It is remarkable that the experimental constraints from $\Delta m_K$ and 
$\epsilon_K$ are avoided for $m_{\tilde q} \gsim 400$GeV with $|(U_{L,R})_{23}| \sim 10^{-2}$ 
and $|(U_{L,R})_{13}| \sim 10^{-3}$.

\section{${\rm Br}(b \to s\gamma)$}

The SUSY contribution to the $b-s$ transition is constrained from the inclusive $b \to s\gamma$ 
decay. In fact, the experimental result~\cite{Barberio:2008fa} agrees well with the SM 
prediction~\cite{Misiak:2006zs}, restricting the extra contribution apart from the SM in the range,
\begin{eqnarray}
 -3 \times 10^{-5} < \Delta {\rm Br}(b \to s\gamma) < 1.4 \times 10^{-4}
 \label{eq:bsg-bound}
\end{eqnarray}
at the 2$\sigma$ level.

\begin{figure}[h]
\begin{center}
\includegraphics[scale=1]{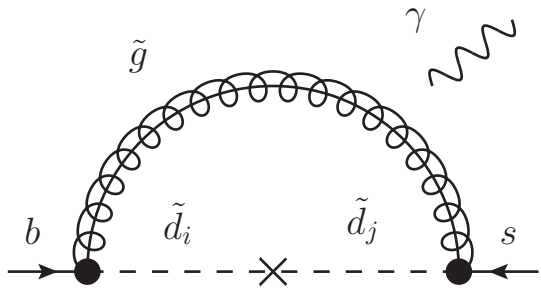} \\ 
\end{center}
\caption{The gluino contribution to $b \to s\gamma$. The cross in the squark propagator 
denotes the chirality flip. The gluino vertex and the chirality flip of the squark have flavor 
structures. }
\label{fig:bsg}
\end{figure}

The $\Delta B = 1$ effective Hamiltonian relevant for ${\rm Br}(b \to s\gamma)$ is
\begin{eqnarray}
 H_{\rm eff} = -\frac{4G_F}{\sqrt{2}} V_{tb} V_{ts} 
 \sum_{i=7\gamma, 8G} C_i O_i + \tilde C_i \tilde O_i,
\end{eqnarray}
where the operators are
\begin{eqnarray}
 O_{7\gamma} = \frac{e}{16\pi^2} m_b \bar s_L (F\cdot\sigma) b_R,~~~
 O_{8G} = \frac{g_s}{16\pi^2} m_b \bar s_L (G\cdot\sigma) b_R,~~~
\end{eqnarray}
and $R \leftrightarrow L$ for $\tilde O_i$. The SUSY contribution to $C_{7\gamma}$ and 
$C_{8G}$ are dominated by the 
gluino diagrams. Furthermore, since the external quarks have opposite chirality to each 
others, the contribution is enhanced by $\tan\beta$ when the chirality flip takes place on 
the down-type squark propagator (see Fig.~\ref{fig:bsg}).
In the hybrid basis, the flavor changing couplings appear in the gaugino vertex and 
the chirality flip of the squark, and the Wilson coefficients become
\begin{eqnarray}
 \label{eq:bsg}
 && C_{7\gamma} = 
 -\frac{\sqrt{2} \alpha_s \pi}{6 G_F V_{tb} V_{ts}} \sum_{i,j} \frac{(m_d)_{ji}}{m_b}
 \frac{m_{\tilde g} \mu \tan\beta}{m_{\tilde qi}^2 m_{\tilde qj}^2} 
 (U_R)_{i3} (U_L^*)_{j2}
 \left[ -\frac{4}{3} M_1(x_i, x_j) \right], \\
 && C_{8G} = 
 -\frac{\sqrt{2} \alpha_s \pi}{2 G_F V_{tb} V_{ts}} \sum_{i,j} \frac{(m_d)_{ji}}{m_b}
 \frac{m_{\tilde g} \mu \tan\beta}{m_{\tilde qi}^2 m_{\tilde qj}^2}
 (U_R)_{i3} (U_L^*)_{j2}
 \left[ -\frac{1}{6} M_1(x_i, x_j) + \frac{3}{2} M_2(x_i, x_j) \right],
 \nonumber
\end{eqnarray}
where $(m_d)_{ij}$ is defined as $(m_d)_{ij} = (Y_d)_{ij} v = (U_L Y_d^{\rm (diag)} 
U_R^\dagger)_{ij} v$. The loop functions are
\begin{eqnarray}
 && M_1(x, y) = 
 \frac{x^2(1-3y)-y^2(1-3x)+x-y}{(x-1)^2(y-1)^2(x-y)}
 - 2\frac{x^2(y-1)^3\ln x-y^2(x-1)^3\ln y}{(x-1)^3(y-1)^3(x-y)},
 \nonumber\\ 
 && M_2(x, y) = 
 \frac{x^2(1+y)-y^2(1+x)-3(x-y)}{(x-1)^2(y-1)^2(x-y)}
 + 2\frac{x(y-1)^3\ln x-y(x-1)^3\ln y}{(x-1)^3(y-1)^3(x-y)},
 \nonumber
\end{eqnarray}
with $x_i = m_{\tilde g}^2/m_{\tilde qi}^2$. Although there are additional contributions from 
chargino, neutralino and charged Higgs, we discard them for simplicity. 

We notice that the split generation model easily satisfies the $b \to s\gamma$ bound in the mass 
region which is indicated by the di-muon anomaly. This is because the bottom-squark contribution 
is suppressed by a large mass, $M_{\tilde q}$. The strange-squark diagram is independent of 
$M_{\tilde q}$, but the contribution is small because the chirality flip is given by $m_s$. 
Consequently, the SUSY contribution to $b \to s\gamma$ is estimated to be $< O(10^{-5})$ for 
$m_{\tilde q} \sim 100$GeV and $M_{\tilde q} \sim 1$TeV with $\tan\beta = 10$ and $|(U_L)_{23}| 
= |(U_R)_{23}| = \lambda^2$. This result is much less than the experimental bound (\ref{eq:bsg-bound}). 
Furthermore, although a large CP violation in the $b-s$ transition can affect the time-dependent 
CP asymmetry of the $B_d \to \phi K$ and $\eta' K$ decays, a contribution from the chromo dipole 
operator, which usually dominates the SUSY contributions (see e.g.~\cite{Endo:2004dc}), 
is suppressed by $M_{\tilde q}$, too. 

The above result is contrasted to other mass models. The SUSY contributions to $b \to s\gamma$ is 
generally large in the degenerate and decoupling mass models, since the third generation is 
relatively light. In fact, the bottom-squark exchange diagram dominates the SUSY contributions. 
Compared with the dominant contribution in the split generation model, the bottom-squark contribution 
is enhanced by $m_b/m_s$ due to a large chirality flip. On the other hand, the $B_s - \bar B_s$ 
oscillation receives almost the same contribution as (\ref{eq:b-bbar}) from the bottom squark. 
Thus, they almost saturates or partially exceeds the experimental constraint. In addition, 
the time-dependent CP asymmetries of $B_d \to \phi K$ and $\eta' K$ are likely to receive a large 
contribution from the bottom-squark diagrams, which may exceed the experimental values. 
The above difficulty is also noted in the analysis on the di-muon anomaly and $b \to s\gamma$ in 
\cite{Ko:2010mn} for the degenerate case, where the squark masses are taken quasi universal 
with respect to the generations, $\tilde m_q \sim {\rm diag}\,(m_{\tilde q}, m_{\tilde q}, m_{\tilde q})$. 
We also obtain a very similar result for the decoupling model, since the bottom-squark contributions 
are not suppressed. 

\section{Electric Dipole Moments}

The anomalous like-sign di-muon charge asymmetry implies that the CP symmetry is 
violated in the squark sector. However, the violations are tightly limited by the EDM experiments. 
We assume throughout the present paper that all CP violating phases come only from the mixing 
matrices for quarks and leptons, as discussed in the Introduction. We see that even if they 
appear from the quark mixings, the electric and chromo-electric dipole moments (EDM's and CEDM's) 
can receive large contributions from superparticle exchanges. 
The CEDM operators are defined by the effective Lagrangian,
\begin{eqnarray}
 \mathcal{L}_{\rm eff} = -\sum_f \frac{i}{2} \tilde d_f g_s \bar\psi_f (G\cdot\sigma) 
 \gamma_5 \psi_f,
 \label{eq:EDM-operator}
\end{eqnarray}
and the EDM operators are given similarly. In particular, according to \cite{Hisano:2003iw}, 
the CEDM of the strange quark is severely constrained by the neutron EDM, 
$e|\tilde d_s| \lsim 1 \times 10^{-25}\,e$cm, with updating the experimental data~\cite{Baker:2006ts}. 
Since the estimation relies on the hadronic calculation and potentially includes a large uncertainty, 
we take a more conservative bound, $e|\tilde d_s| < O(10^{-(24-25)})e$cm, in the following study. 

Through the $2-3$ generation mixings, the Yukawa coupling receives a large contribution 
with the bottom quark mass, $(Y_d)_{22} \supset (Y_d^{\rm (diag)})_{33} (U_{L})_{23} 
(U_{R}^*)_{23}$. Since the mixings $(U_{L})_{23}$ and $(U_{R})_{23}$ are complex in general, 
we obtain the SUSY contribution to the CEDM from the strange-squark loop as
(c.f. \cite{Hisano:2003iw})
\begin{eqnarray}
 \tilde d_s = 
 \frac{\alpha_s}{4\pi} 
 \frac{m_b m_{\tilde g} \mu\tan\beta}{m_{\tilde q}^4}
 |(U_{L})_{23}(U_{R})_{23}^*| \sin\Theta^-
 \left( -\frac{1}{6} M_1(x) + \frac{3}{2} M_2(x) \right),
 \label{eq:EDM}
\end{eqnarray}
where the CP phase is
\begin{eqnarray}
  \Theta^- \equiv \arg[(U_{L})_{23}(U_{R})_{23}^*].
\end{eqnarray}
It is noticed that $\Theta^-$ is different from $\Theta^+$ in (\ref{eq:phase+}). Also, we find 
that the contribution is independent of the heavy squark mass, $M_{\tilde q}$. Here, the 
loop functions are defined as
\begin{eqnarray}
 M_1(x) = \frac{1+4x-5x^2+2(2+x)x\ln x}{(x-1)^4},~~
 M_2(x) = \frac{5-4x-x^2+2(1+2x)\ln x}{(x-1)^4},
\end{eqnarray}
with $x = m_{\tilde g}^2/m_{\tilde q}^2$. The bottom squark gives the contribution 
as Eq.~(\ref{eq:EDM}) with substituting the squark mass $m_{\tilde q}$ by $M_{\tilde q}$,
which is suppressed roughly by $1/M_{\tilde q}^4$.

The SUSY contribution (\ref{eq:EDM}) easily exceeds the experimental bound. 
We obtain a constraint on the CP phase $\Theta^-$ as
\begin{eqnarray}
  |\sin\Theta^-| \lsim 0.1(0.01)\,
  \left(\frac{\tan\beta}{10}\right)^{-1}
  \left(\frac{m_{\tilde q}}{500{\rm GeV}}\right)^{2},
\end{eqnarray}
from the experimental constraint $e|\tilde d_s| < 10^{-24} (10^{-25})e$cm for $|(U_{L})_{23}| 
= |(U_{R})_{23}| = \lambda^2$ with assuming $m_{\tilde q} = m_{\tilde g} = \mu$. Thus, a 
phase of $(U_{L})_{23}$ is required to coincide with that of $(U_{R})_{23}$ at the $10(1)$\% 
accuracy. 

In spite of the tight constraint on the phases, the CP phase of the $B_s - \bar B_s$ oscillation 
can be large. This is because $\Theta^-$ depends on $(U_{L})_{23}$ and $(U_{R})_{23}$ 
differently from $\Theta^+$ which appears in $B_s - \bar B_s$. Namely, the $B_s - \bar B_s$ 
mixing receives a contribution from $\arg(U_{L})_{23} + \arg(U_{R})_{23}$, while the EDM depends 
on $\arg(U_{L})_{23} - \arg(U_{R})_{23}$. Thus, an $O(1)$ phase is allowed for $(U_{L})_{23}$ 
and $(U_{R})_{23}$, as long as it satisfies $\arg(U_{L})_{23} \simeq \arg(U_{R})_{23}$. Such a 
condition is naturally realized when the Yukawa coupling is hermitian, since the rotation matrices 
become $U_L = U_R$, and the CEDM is suppressed, $\Theta^- = 0$. This hypothesis of the 
hermit Yukawa couplings may be very interesting in a connection with a solution to the strong 
CP problem~\cite{Hiller:2002um}, as discussed in the Introduction. 

So far, we have discussed the case when both $(U_{L})_{23}$ and $(U_{R})_{23}$ are large. 
If either of them is suppressed, the situation gets worse. Since the SUSY contribution to the 
$B_s - \bar B_s$ mixing dumps rapidly as is found from (\ref{eq:b-bbar}), we need a larger 
mixing angle to explain the di-muon anomaly such as $|(U_{L})_{23}|$ or $|(U_{R})_{23}| \sim 1$ 
for $\bar m_{\tilde q} = 500$GeV. Then, we obtain severer constraints from $b \to s\gamma$ 
and/or the chargino or two-loop contributions to the neutron/atomic 
EDM's~\cite{Endo:2003te,Hisano:2008hn}. 

\section{Lepton Flavor Violations}

Let us discuss flavor violations in the lepton sector. A generation mass splitting of the sleptons 
generally induces the lepton flavor violations (LFV's) through mixings between the leptons 
similarly to the squarks. In particular, if we consider the grand unification (GUT), 
the lepton sector is related to the quark sector, and we expect that the mixing angles of the lepton 
Yukawa couplings are also given by the CKM matrix. In this section, we analyze the flavor violating 
decay of the muon and the tau leptons in the hybrid flavor basis. 

The severest constraint on the LFV's comes from the $\mu \to e \gamma$ decay. From the 
experiments, the bound is known to be ${\rm Br}(\mu \to e\gamma) < 1.2 \times 10^{-11}$ 
at 90\%~\cite{PDG}. The smuon--selectron diagram with the flavor-changing Yukawa coupling 
gives the branching ratio as
\begin{eqnarray}
 {\rm Br}(\mu \to e\gamma) 
 &\simeq&
 \frac{1}{\Gamma_{\rm tot}}
 \frac{\alpha \alpha^{\prime 2}}{1152\pi^2} m_\mu^3 m_\tau^2 \epsilon^2 
 \frac{M_1^2 \mu^2 \tan^2\beta}{m_{\tilde \ell}^8}
 \nonumber \\
 &\simeq&
 2 \times 10^{-12} 
 \left(\frac{\epsilon}{10^{-5}}\right)^{2}
 \left(\frac{\tan\beta}{10}\right)^{2}
 \left(\frac{m_{\tilde\ell}}{400{\rm GeV}}\right)^{-6}
 \left(\frac{\mu}{1{\rm TeV}}\right)^{2},
\end{eqnarray}
in the limit of decoupled heavy sleptons, $M_{\tilde\ell} \gg m_{\tilde\ell}$. 
Here, $\Gamma_{\rm tot}$ is the total decay rate of the muon, and we assumed the Bino mass 
$M_1$ is equal to the light slepton mass $m_{\tilde \ell}$, for simplicity. 
The mixing angle $\epsilon$ is given by the lepton diagonalization 
matrices, $U_{\ell_{L, R}}$, such as $(U_{\ell_L})_{23}(U_{\ell_R}^*)_{13}$ and $(U_{\ell_R})_{23}
(U_{\ell_L}^*)_{13}$. Postulating $(U_{\ell_L})_{23} \sim (U_{\ell_R})_{23} \sim 10^{-2}$ and 
$(U_{\ell_L})_{13} \sim (U_{\ell_R})_{13} \sim 10^{-3}$ as is expected from the CKM matrix, 
the mixing angle is roughly $\epsilon \sim 10^{-5}$. Compared with the experimental 
result~\cite{PDG}, we find that the split generation model satisfies the current constraint. 
Furthermore, it is in the reach of the sensitivity of the MEG experiment~\cite{Ritt:2006cg}, 
as long as $\mu\tan\beta$ is not very small. 

Similarly to the $b-s$ transition, the tau lepton can decay into the muon. The branching ratio 
of $\tau \to \mu\gamma$ is bounded by the experiments, ${\rm Br}(\tau \to \mu\gamma) 
< 4.4 \times 10^{-8}$ at 90\%~\cite{PDG}. Among the SUSY contributions, the smuon--chargino 
diagram is dominant, which is evaluated as
\begin{eqnarray}
 {\rm Br}(\tau\to\mu\gamma) 
 &\simeq&
 \frac{1}{\Gamma_{\rm tot}}
 \frac{\alpha \alpha_2^2}{144\pi^2} m_{\tau}^5 
 \frac{M_2^2 \tan^2\beta}{\mu^2 m_{\tilde\ell}^4} |(U_{\ell_L})_{23}|^2
 \nonumber \\
 &\simeq&
 1 \times 10^{-8} 
 \left(\frac{|(U_{\ell_L})_{23}|}{\lambda^2}\right)^{2}
 \left(\frac{\tan\beta}{10}\right)^{2}
 \left(\frac{m_{\tilde\ell}}{400{\rm GeV}}\right)^{-2}
 \left(\frac{\mu}{2{\rm TeV}}\right)^{-2},
\end{eqnarray}
in the decoupling limit, where $\Gamma_{\rm tot}$ means the total decay rate of the tau lepton, 
and $M_2 = m_{\tilde\ell}$ is assumed in the second line. We notice that the rate is suppressed 
by a large $\mu$ parameter, since the diagram includes the mixed Wino- and Higgsino-like 
charginos. Although the pure Wino contribution is not suppressed by $M_{\tilde\ell}$, this 
results in ${\rm Br}(\tau\to\mu\gamma) \simeq 10^{-11}$ for $m_{\tilde \ell} = 400$GeV. 
As a result, we found that the SUSY contribution to $\tau\to\mu\gamma$ easily satisfies the current 
experimental bound~\cite{PDG}. Furthermore, if the Higgsino is relatively light and $\tan\beta$ 
is large, the branching ratio may be accessible in the super B factory~\cite{Collaboration:2010af}.

\section{Conclusions}

In this letter, we have studied the split generation model, where the scalar fermions in the first and second 
generations have small and degenerate masses of $O(100)$GeV, and those in the third generation 
are heavy with mass of $O(1-10)$TeV. The severe constraints from the FCNC's such as the 
$K - \bar K$ mixing are suppressed sufficiently, while the mass difference between the 
squark generations generally induces the $b-s$ transitions with a new CP phase. We have found that the 
SUSY contribution to the $B_s - \bar B_s$ mixing can be large enough to explain the current anomaly 
of the like-sign di-muon charge asymmetry in the D0 experiment. We have also discussed that the 
contribution is consistent with the other experimental constraints such as $b \to s\gamma$ and EDM's. 
This situation is contrasted to other mass models such as the degenerate or the decoupling models. 

The model is also interesting from other view points of phenomenology. In addition to the 
di-muon anomaly, it generally predicts flavor violations in the lepton sector. It has been shown that 
the branching ratios of $\mu \to e\gamma$ and $\tau \to \mu\gamma$ can be close to the sensitivities 
of the future experiments. Furthermore, the heavy third generation enhances light lepton events 
in the LHC experiment, which improves the SUSY detections and measurements~\cite{MNSY}. 
Also, a light mass of the second generations allows a large contribution to the muon anomalous 
magnetic moment, which currently has a $3-4\sigma$ discrepancy between the experimental data 
and the SM prediction~\cite{Hagiwara:2006jt}. Lastly, a heavy third generation is favored to satisfy 
the mass bound on the neutral Higgs boson from the LEP experiment~\cite{PDG}. We expect to 
test the split generation model in future experiments such as the LHC, LHCb, EDM's, 
superB factories, and MEG.

\end{document}